\documentclass[preprint,showpacs,preprintnumbers,amsmath,amssymb]{revtex4}

\usepackage{graphicx}
\usepackage{epsfig}
\usepackage{bm}
\usepackage{amsfonts}
\usepackage{subfigure}
\usepackage{multirow}
\usepackage{array}
\usepackage{diagbox}
\usepackage{makecell}
\usepackage{slashbox}

\begin{document}



\title[]{Resurrecting the exponential and inverse power-law potentials in non-canonical inflation}

\author{Zeinab Teimoori\footnote{zteimoori16@gmail.com} and Kayoomars Karami\footnote{kkarami@uok.ac.ir}}

\address{Department of Physics, University of Kurdistan, Pasdaran
Street, P.O. Box 66177-15175, Sanandaj, Iran}

\begin{abstract}
We study inflation within the framework of non-canonical scalar field. In this scenario, we obtain the inflationary observables such as the scalar spectral index, the tensor-to-scalar ratio, the running of the scalar spectral index as well as the equilateral non-Gaussianity parameter. Then, we apply these results for the exponential and inverse power-law potentials. Our investigation shows that although the predictions of these potentials in the standard canonical inflation are completely ruled out by the Planck 2015 observations, their results in non-canonical scenario can lie inside the allowed regions of the Planck 2015 data. We also find that in non-canonical inflation, the predictions of the aforementioned potentials for the running of the scalar spectral index and the equilateral non-Gaussianity parameter are in well agreement with the Planck 2015 results. Furthermore, we show that in the context of non-canonical inflation, the graceful exit problem of the exponential and inverse power-law potentials is resolved.
\end{abstract}



\maketitle


\section{Introduction}\label{sec:int}

Standard Big Bang cosmology is an extremely successful model at describing considerable issues such as the Cosmic Microwave Background (CMB) and the light nucleosynthesis. Despite these remarkable success, it suffers from serious problems such as the flatness problem, the horizon problem and the magnetic monopole problem. Inflation is a beautiful idea which was first proposed in the early 1980s and solved all of these problems \cite{Starobinsky1980, Kazanas1980, Sato1981a, Sato1981, Guth1981, Linde1982, Albrecht1982, Linde1983}. In addition, inflation explains the origin of the anisotropy seen in the CMB radiation as well as the Large Scale Structure (LSS) formation in the universe \cite{Mukhanov1981, Hawking1982, Starobinsky1982, Guth1982}. According to the inflation theory, our universe undergoes an accelerated expansion period in the early stages of its evolution \cite{Liddle2000,Lemoine2000}.
The standard inflationary scenario is based on a canonical scalar field (called inflaton) which is minimally coupled to the Einstein gravity.

The non-canonical Lagrangian densities, motivated by string theory have also been used in the inflationary models. In this class of models, the kinetic term is different relative to the canonical case \cite{Armendariz1999, Garriga1999, Chiba2000, Armendariz2001, Unnik2012, Unnik2013, Rezazade15, Felice2011, Tsujikawa2013}. One class of non-canonical models is k-inflation in which the kinetic term can dominate the potential one \cite{Armendariz1999,Garriga1999,Chiba2000,Armendariz2001}. In \cite{Unnik2012}, the authors studied various inflationary potentials and showed that in the non-canonical scenario, the viability of inflationary models can be improved.

There is one important class of non-canonical models in which the non-canonical kinetic term is shown by a general form as $\omega(\phi) X$, where $X$ is the canonical kinetic term and $\omega(\phi)$ is a general function of the scalar field $\phi$. The authors of \cite{Felice2011} considered the non-canonical version of the chaotic inflationary potential, $V(\phi)=V_0\left(\phi/M_{\rm pl}\right)^{p}$, with the exponential coupling $\omega(\phi)= e^{\mu (\phi/M_{\rm pl})}$ and obtained the observational constraints on this model. They showed that in contrary to the standard canonical scenario, the results of the chaotic potentials in the non-canonical framework are compatible with the observations deduced from the WMAP7 data (see also \cite{Tsujikawa2013}). This class of non-canonical models can be considered as a subset of the Horndeski theory \cite{Horndeski1974, Deffayet2011, Charmousis2012, Kobayashi2011, Felice2011,Tsujikawa2013}. The Lagrangian of the
Horndeski theory encompasses a wide range of gravitational theories, such as standard canonical inflation\cite{Linde1982,Linde1983}, non-minimally coupled models \cite{Fakir1990,Bezrukov2008}, non-canonical scalar field models \cite{Nakayama2010, Felice2011}, Brans-Dicke theories \cite{Brans1961}, Galileon inflation \cite{Kobayashi2010}, field derivative
couplings to gravity \cite{Amendola1993} and k-inflation \cite{Armendariz1999,Garriga1999}.

  In the present work, we focus on the non-canonical scalar field models and examine the viability of the two important inflationary potentials including the exponential and inverse power-law potentials in light of the Planck 2015 data. The predictions of these potentials in the standard canonical inflation are incompatible with the observations \cite{Rezazade15,Planck2015,Rezazade2015,Rezazade2016}. It is interesting to see whether the non-canonical versions of the aforementioned potentials can be resurrected in light of the Planck 2015 results. Furthermore, the exponential and inverse power-law potentials in the standard canonical inflation suffer from a serious problem known as the ``graceful exit'' problem in which inflation never ends and continues forever \cite{Unnik2013,Bar93}. We try to investigate whether this problem can be solved in the framework of non-canonical inflation.

  This paper is organized as follows. In Sec. \ref{review:sec}, we review the non-canonical inflationary scenario and obtain the necessary relations governing the inflationary observables. In Secs. \ref{sec:exp} and \ref{sec:inverse}, we apply the results of Sec. \ref{review:sec} for the exponential potential and the inverse power-law potential, respectively, and investigate the viability of these models in light of the Planck 2015 data. Our conclusions are presented in Sec. \ref{sec:con}.

\section{Inflation with non-canonical scalar field}\label{review:sec}

The action of a scalar field with the non-canonical kinetic term $\omega(\phi)X$ and the potential $V(\phi)$ is given by \cite{Felice2011,Tsujikawa2013}
\begin{equation}\label{action}
S= \int d^{4}x \sqrt{-g}\left[\frac{M_{\rm pl}^2}{2} R + \omega(\phi)X -V(\phi) \right],
\end{equation}
where the scalar field $\phi$ is minimally coupled to the Einstein gravity. Here, $M_{\rm pl}=(8\pi G)^{-1/2}$ is the reduced Planck mass, $g$ is the determinant of the metric $ g_{{\mu}{\nu}}$, $R$ is the Ricci scalar and $\omega (\phi)$ is the field coupling with the kinetic term $X\equiv -\frac{1}{2}g^{\mu\nu}\phi_{,\mu}\phi_{,\nu}$.

Taking the variation of the action (\ref{action}) with respect
to the flat Friedmann-Robertson-Walker (FRW) metric $ g_{{\mu}{\nu}}={\rm diag}\left(-1,a^{2}(t),a^{2}(t),a^{2}(t)\right)$, one can obtain the modified Friedmann equations as
\begin{align}
\label{FR1:eq}
  3 M_{\rm pl}^{2} H^{2}-\omega(\phi) X -V(\phi) &=0\,, \\
  \label{FR2:eq}
  3 M_{\rm pl}^{2} H^{2}+2 M_{\rm pl}^{2} \dot{H}+\omega(\phi) X -V(\phi)&=0\,,
\end{align}
where $X=\dot{\phi}^2/2$, $H\equiv \dot{a}/a $ is the Hubble parameter and a dot indicates derivative with respect to the cosmic time $t$.

In the flat FRW background, varying the action (\ref{action}) with respect to the scalar field $\phi$ leads to the equation of motion
\begin{equation}\label{Field:eq}
  \omega(\phi) \,\ddot{\phi}+\big(3\,\omega(\phi) H + \dot{\phi}\,\omega_{,\phi}\big)\dot{\phi}-\omega_{,\phi}X +V_{,\phi}=0\,,
\end{equation}
where $({,\phi})\equiv d/d\phi$\,.\\
In the context of non-canonical inflation, the slow-roll parameters are defined as \cite{Felice2011}
\begin{equation}\label{SRP}
\varepsilon \equiv -\frac{\dot H}{H^2}, ‎\hspace{.5cm} \delta_{X}\equiv \frac{\omega X}{M_{\rm pl}^{2} H^2},‎ \hspace{.5cm} \delta_{\phi}\equiv \frac{\ddot{\phi}}{ H\, \dot{\phi}}\, .
\end{equation}
In order to match our model with the observations, we need to calculate the inflationary observable parameters. To finding these parameters it is needed to study the scalar and tensor modes of perturbations during inflation. Using the Arnowitt-Deser-Misner (ADM) formalism, one can obtain the power spectrum of the scalar (curvature) perturbations in the slow roll limit as \cite{Felice2011,Felice11}
\begin{equation}\label{Ps}
{\cal P}_{s}\simeq\frac{H^2}{8 \pi ^{2} Q_{s}}\,,
\end{equation}
where
\begin{equation}\label{Qs}
Q_{s}\equiv\frac{\omega(\phi) {\dot{\phi}}^2}{2 H^2}\,.
\end{equation}

One of the important inflationary observables is the scalar spectral index $n_s$ which determines the scale dependence of the scalar perturbation spectrum. This observable is defined as
\begin{equation}\label{ns1}
n_{s}-1\equiv \frac{d\ln{{\cal P}_{s}}}{d\ln{k}}\Big|_{k=aH},
\end{equation}
where the calculation should be done at the time of horizon exit, i.e. $k=aH$. The scalar spectral index reported by the Planck team is about $ n_{s}=0.9644 \pm 0.0049 $ (68\% CL, Planck 2015 TT, TE, EE+lowP) \cite{Planck2015}.

Using the scalar power spectrum (\ref{Ps}) and the approximation $d\ln k \simeq H dt$, the scalar spectral index (\ref{ns1}) at first order in slow roll parameters takes the form \begin{equation}\label{ns}
n_{s}-1\simeq {-4 \varepsilon}+2\eta_{H} \,,
\end{equation}
where  $\eta_{H}$ is defined as
\begin{equation}\label{eta:H}
\eta_{H}\equiv \frac{-{\ddot H}}{2 H {\dot H}} = \varepsilon-\frac{\eta}{2}\,,
\end{equation}
and
\begin{equation}\label{eta}
\eta\equiv \frac{\dot{\varepsilon}}{H \varepsilon}.
 \end{equation}

From Eq. (\ref{ns}), one can calculate another observable called the running of the scalar spectral index $d n_{s}/d\ln k$ at second order in slow roll parameters as
\begin{equation}\label{dns}
\frac{d n_{s}}{d\ln k}\simeq \frac{d n_{s}}{Hdt}\simeq-8\,{\varepsilon^{2}}+ 10 \,{\varepsilon}\, {\eta_{H}}+ 4 {\eta_{H}^{2}}+2\, {\xi_{H}}{\eta_{H}},\,
\end{equation}
where
\begin{equation}\label{xiH}
\xi_{H}\equiv\frac{\dddot H}{H {\ddot H}}.
\end{equation}
Observational value for the running of the scalar spectral index is $d{n_s}/d\ln k =  - {\rm{0}}{\rm{.0085}} \pm {\rm{0}}{\rm{.0076}}$ (68\% CL, Planck 2015 TT, TE, EE+lowP) \cite{Planck2015}.

Now we are interested in obtaining the power spectrum of the tensor perturbations. The tensor power spectrum can be calculated in a similar way to the one followed for the scalar modes and in the slow roll regime it reads \cite{Felice2011,Felice11}
\begin{equation}\label{Pt}
{\cal P}_{t}\simeq\frac{2 H^2}{\pi ^{2} M_{\rm pl}^{2}}\,.
\end{equation}
The scale dependence of the tensor power spectrum is determined by the tensor spectral index defined as
\begin{equation}\label{nt1}
n_{t}\equiv \frac{d\ln{{\cal P}_{t}}}{d\ln{k}}\Big|_{k=aH},
\end{equation}
where at first order in terms of the slow roll parameters, it reads
\begin{equation}\label{nt}
n_{t}\simeq -2 \varepsilon.
\end{equation}

Another important observable parameter is the tensor-to-scalar ratio defined as
\begin{equation}\label{r}
r\equiv\frac{{\cal P}_{t}}{{\cal P}_{s}},
\end{equation}
where the Planck 2015 data predicts an upper bound on the tensor-to-scalar ratio as $r<0.149$ (95\% CL, Planck 2015 TT, TE, EE+lowP) \cite{Planck2015}. The $r$ parameter in our model with using Eqs. (\ref{Ps}), (\ref{Pt}) and (\ref{r}) in the slow-roll approximation turns into
\begin{equation}\label{r:epsilon}
r \simeq 16 \frac{Q_{s}}{M_{\rm pl}^{2}}= 16 \epsilon\,.
\end{equation}
 Note that from Eqs. (\ref{nt}) and (\ref{r:epsilon}), one can easily get the consistency relation in non-canonical scenario as $ r = - 8 n_{t}$ which is same as that obtained in the standard canonical inflation.

One another observable predicted by inflation is the primordial non-Gaussianity which can be generated due to quantum mechanical effects at or before horizon exit. The primordial non-Gaussianity can be used as a powerful tool to distinguish between different inflationary models. Single field inflationary models with non-canonical kinetic term generate the equilateral form of the non-Gaussianity, so in this work we focus on the equilateral configuration. The equilateral non-Gaussianity $ f_{\rm NL}^{\rm equil}$, in the slow-roll approximation has been evaluated in \cite{Felice2011,Felice11} as
\begin{equation}\label{fNL}
f_{\rm NL}^{\rm equil}\simeq \frac{85}{36}\,\varepsilon - \frac{5}{6}\,\eta_{H}\,.
\end{equation}
Equation (\ref{fNL}) indicates the equilateral non-Gaussianity is of order of the slow-roll parameters. Therefore, we can claim that our model in the slow-roll regime can produce small non-Gaussianity. The measured value of the equilateral non-Gaussianity is $f_{{\rm{NL}}}^{{\rm{equil}}} =  - 4 \pm 43$ (68\% CL, Planck 2015 T+E) \cite{Planck2015}.

So far, we have obtained the inflationary observables $n_{s}$, $r$, $\frac{d n_{s}}{d\ln k} $ and $f_{\rm NL}^{\rm equil}$ in terms of the slow-roll parameters. However, it is convenient to find them in terms of the potential $V(\phi)$ and the field coupling $\omega(\phi)$ and in this regards the parameter $\varepsilon \equiv -\dot{H}/H^2$ can be expresses as \cite{Felice2011}
\begin{equation}\label{epsilon:v}
\varepsilon\simeq \frac{M_{\rm pl}^2}{2} \frac{1}{B^2}\left(\frac{V_{,\phi}}{V}\right)^2\,,
\end{equation}
where $ B(\phi)\equiv \omega(\phi)^{1/2}$.
Using this, Eqs. (\ref{ns}) and (\ref{r:epsilon}) can be rewritten in the following forms
\begin{align}
  \label{ns:V}
  n_{s}-1 & \simeq \frac{M_{\rm pl}^2}{B^2}\left[ 2 \, \frac{V_{,\phi \phi}}{V}-3 \,\frac{{V_{,\phi}}^{2}}{V^2}-2\, \frac{B_{,\phi}}{B}\,\frac{V_{,\phi}}{V}\right],\, \\
   \label{r:V}
  r & \simeq 8 \,\frac{M_{\rm pl}^2}{B^2}\,\frac{V_{,\phi}^2}{V^2}\,.
\end{align}
In a similar way, the running of the scalar spectral index (\ref{dns}) and the equilateral non-Gaussianity parameter (\ref{fNL}) turn into
\begin{eqnarray}  \label{dns:V}
   \frac{d n_{s}}{d\ln k} \simeq
   2\,\frac{M_{\rm pl}^4}{B^4}\,\frac{V_{,\phi}}{V}\left[-\frac{V_{,\phi \phi \phi}}{V}-3\,\frac{V_{,\phi}^3}{V^3}-3\,\frac{B_{,\phi}^2}{B^2}\frac{V_{,\phi}}{V}+\frac{V_{,\phi}}{V}\left(\frac{B_{,\phi \phi}}{B}+4\,\frac{V_{,\phi \phi}}{V}\right)\right.~~~\nonumber\\+\left.
    \frac{B_{,\phi}}{B}\left(3\,\frac{V_{,\phi \phi}}{V}-4\,\frac{V_{,\phi}^2}{V^2}\right)\right],
\end{eqnarray}
\begin{eqnarray}\label{fNL:V}
 f_{\rm NL}^{\rm equil}& \simeq &
  \frac{5}{72}\,\frac{M_{\rm pl}^2}{B^2}\left[23\,\frac{V_{,\phi}^2}{V^2} -12\,\frac{V_{,\phi \phi}}{V}+ 12\,\frac{B_{,\phi}}{B}\frac{V_{,\phi}}{V} \right].
 \end{eqnarray}

At the end of this section, we introduce the $e$-fold number which describes the amount of inflation between two times, $t$ and $t_{f}$, and is defined as
\begin{equation}\label{N}
N \equiv\int_{t}^{t_f} H dt,
\end{equation}
where $t_{f}$ is the ending time of inflation. Note that around $40$ to $60$ $e$-folds are required to solve the flatness and horizon problems \cite{Randall1996,Barger2003}. The number of $e$-folds $N$ in the non-canonical scenario under the slow-roll approximation is given by \cite{Felice2011}
\begin{equation}\label{N:v}
N \simeq \frac{1}{M_{\rm pl}}\int_{x_f}^{x} B^2(\phi)\left(\frac{V}{V_{,\phi}}\right) dx\,,
\end{equation}
where $x\equiv \frac{\phi}{M_{\rm pl}}$, $x_{f}\equiv \frac{\phi_{f}}{M_{\rm pl}}$ and $\phi_{f}$ is the field value at the end of inflation.

Until now, we have obtained the essential relations governing the inflationary observables containing $n_{s}$, $r$, $\frac{d n_{s}}{d\ln k}$ and $f_{{\rm{NL}}}^{{\rm{equil}}}$ in the non-canonical inflationary model described by the action (\ref{action}). In what follows, we apply these results for the exponential and inverse power-law potentials to check their viability in light of the Planck 2015 results. Note that the results of the aforementioned potentials in the standard canonical inflationary setting are completely ruled out by the Planck 2015 data \cite{Rezazade15,Planck2015,Rezazade2015,Rezazade2016}. Besides, we are also interested in investigating how the graceful exit problem of the exponential and inverse power-law potentials can be solved in our non-canonical model.

\section{exponential potential in non-canonical inflation}\label{sec:exp}

First, we examine one of interesting inflationary potentials which has an exponential form as
\begin{equation}\label{v:exp}
V(\phi)=V_{0}\,{e^{-\alpha \,(\phi/M_{\rm pl})}}\,,
\end{equation}
where $V_{0}$ and $\alpha>0$ are constants. In the standard canonical inflation, the exponential potential (\ref{v:exp}) leads to the power-law inflation with the scale factor $a(t)\propto t^{q}$ where $q>1$ \cite{Liddle2000,Luc85,Hall87,Yok88,Mar14}. The power-law inflation in the canonical scenario suffers from the graceful exit problem in which inflation never ends \cite{Unnik2013}. Furthermore, in the standard canonical setting, the exponential potential (\ref{v:exp}) is disfavored by the Planck 2015 data \cite{Planck2015,Rezazade2015,Rezazade2016}. All these motivate us to examine the potential (\ref{v:exp}) in the non-canonical inflationary scenario to see whether this potential can be resurrected in light of the Planck 2015 results and its graceful exit problem can be solved. To this end, in the action (\ref{action}) we consider the field-dependent coupling term as
\begin{equation}\label{w:exp}
\omega(\phi)= e^{-\mu (\phi/M_{\rm pl})},
\end{equation}
 where $\mu$ is a constant. This coupling is motivated by the dilatonic scalar field in string theory \cite{Felice2011,Gasperini2003}.

For the model described by the potential (\ref{v:exp}) and the coupling term (\ref{w:exp}), using Eqs. (\ref{ns:V}) and (\ref{r:V}), the scalar spectral index $n_{s}$ and the tensor-to-scalar ratio $r$ can be obtained in terms of $x\equiv \phi/M_{\rm pl}$ as
\begin{align}
\label{ns:exp}
 n_{s}-1 &\simeq -\, e^{\mu x}{\alpha ^2}\left(1+\frac{\mu}{\alpha}\right),\\
  \label{r:exp}
  r & \simeq 8 \,e^{\mu x} {\alpha ^2}.
\end{align}
With the help of Eq. (\ref{N:v}), the $e$-fold number reads
\begin{equation}\label{N:exp}
N\simeq \frac{1}{\alpha \mu}\left(e^{-\mu x}- e^{- \mu x_{f}}\right).
\end{equation}
Applying the end of inflation condition $\varepsilon =1$ to Eq. (\ref{epsilon:v}) one can find
\begin{equation}\label{xf:exp}
x_{f}=\frac{1}{\mu}\ln\left(\frac{2}{\alpha^2}\right),
\end{equation}
which is the scalar field at the end of inflation. The result (\ref{xf:exp}) shows that in contrary to the standard canonical inflation, in the non-canonical scenario described by the action (\ref{action}) and with the dilatonic coupling term (\ref{w:exp}), the graceful exit problem of the potential (\ref{v:exp}) is solved.

Now we can substitute Eq. (\ref{xf:exp}) into (\ref{N:exp}) and in this way the $e$-fold number is obtained in terms of the scalar field $x$ as
\begin{equation}\label{N:x}
N\simeq\frac{1}{2\alpha \mu}\left(2\,e^{-\mu x}-{\alpha^2}\right)\,.
\end{equation}
Solving Eq. (\ref{N:x}) for $x$ leads to the following expression
\begin{equation}\label{x:N}
x\simeq\frac{1}{\mu}\ln\left(\frac{2}{\alpha^2(1+2\,N\,\frac{\mu}{\alpha})} \right)\,.
\end{equation}

Finally, using Eq. (\ref{x:N}) in (\ref{ns:exp}) and (\ref{r:exp}) we can rewrite the scalar spectral index $n_{s}$ and the tensor-to-scalar ratio $r$ in terms of the $e$-fold number as
\begin{align}
  \label{ns:fexp}
  n_{s}-1&\simeq \frac{-2 \,(1+\lambda)}{1+2 \,\lambda \,N}\,,  \\
  \label{r:fexp}
  r & \simeq \frac{16}{1+2 \,\lambda \,N}\,,
\end{align}
where $\lambda \equiv \mu/\alpha$. These relations show that the observables $n_s$ and $r$ for the exponential potential (\ref{v:exp}) and the coupling term (\ref{w:exp}) in the non-canonical scenario depend only on the parameter $\lambda$ and the $e$-fold number $N$. Furthermore, combining Eqs. (\ref{ns:fexp}) and (\ref{r:fexp}) yields the linear relation
\begin{equation}\label{rns:exp}
r\simeq \frac{8\,(1-n_{s})}{1+\lambda}.
\end{equation}
We see that if $ \lambda=0$, i.e. $\omega(\phi)=1$, Eq. (\ref{rns:exp}) reduces to $r\simeq 8\,(1-n_{s})$ which is the same result obtained for the power-law inflation in the standard canonical framework \cite{Mar14}.

In the following, we try to find the running of the scalar spectral index $\frac{d n_{s}}{d\ln k}$ and the equilateral non-Gaussianity $f_{\rm NL}^{\rm equil}$ in terms of the $e$-fold number $N$. To do so, at first we use Eqs. (\ref{v:exp}) and (\ref{w:exp}) in Eqs. (\ref{dns:V})-(\ref{fNL:V}) and obtain the following expressions for $\frac{d n_{s}}{d\ln k}$ and $f_{\rm NL}^{\rm equil}$ in terms of the scalar field $x$ as
 \begin{align}
  \label{dns:x:exp}
 \frac{d n_{s}}{d\ln k} &\simeq -\mu \left(1 + \frac{\mu}{\alpha}\right)\alpha ^{3}{e^{2 \mu x}},  \\
  \label{fNL:x:exp}
 f_{\rm NL}^{\rm equil} & \simeq \frac{5}{72}\left(11 + 6 \, \frac{\mu}{\alpha}\right)\alpha ^{2}{e^{ \mu x}}.
\end{align}
Then, replacing $x$ from Eq. (\ref{x:N}) into (\ref{dns:x:exp}) and (\ref{fNL:x:exp}) we get
\begin{align}
  \label{dns:f:exp}
 \frac{d n_{s}}{d\ln k} &\simeq -\frac{4\,\lambda \,(1 +\lambda)}{(1+2\lambda N)^2}\,,  \\
  \label{fNL:f:exp}
 f_{\rm NL}^{\rm equil} & \simeq \frac{5}{36}\left(\frac{11+6 \lambda}{1+2\lambda N}\right).
\end{align}

\begin{figure}[t]
\begin{center}
\scalebox{0.95}[0.95]{\includegraphics{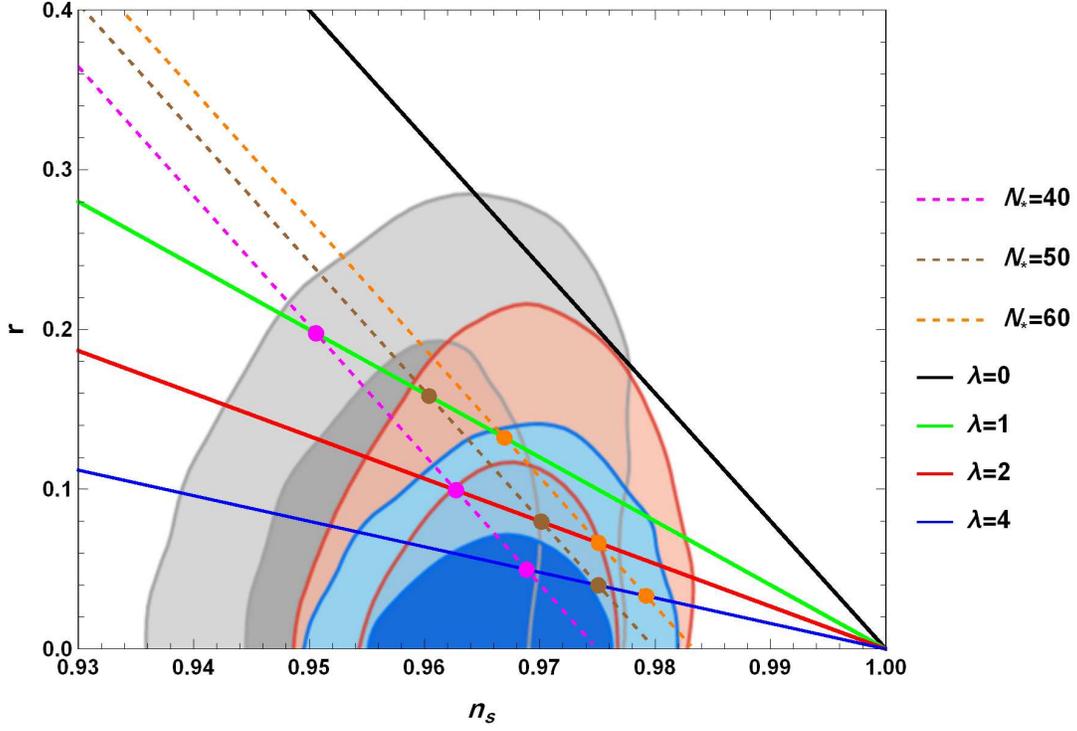}}
\caption{The $r-n_{s}$ diagram for the exponential potential $V(\phi)=V_{0}\,{e^{-\alpha \,(\phi/M_{\rm pl})}}$ with the dilatonic coupling $\omega(\phi)=e^{-\mu (\phi/M_{\rm pl})}$. The results for $N_{*}=$40, 50 and 60 are shown by the pink, brown and orange dashed lines, respectively. Also, the results for $\lambda=$ 0, 1, 2, and 4 are shown by the black, green, red and blue solid lines, respectively. The case of $\lambda=0$ corresponds to the standard canonical inflation. The marginalized joint 68\% and 95\% CL regions of Planck 2013, Planck 2015 TT+lowP and Planck 2015
TT, TE, EE+lowP data \cite{Planck2015} are specified by gray, red and blue, respectively.}
\label{fig:dq}
\end{center}
\end{figure}

Now we turn to Eqs. (\ref{ns:fexp}) and (\ref{r:fexp}) and calculate the observables $n_s$ and $r$. We plot the results in $r-n_s$ plane in Fig. \ref{fig:dq} and compare the prediction of the model with the Planck data. In this figure, the pink, brown and orange dashed lines are corresponding to $N_{*}=$ 40, 50 and 60, respectively, with varying $\lambda$. Also, using Eq. (\ref{rns:exp}) the results for different values of $\lambda=$ 0, 1, 2 and 4 are characterized by the black, green, red and blue solid lines, respectively. In Fig. \ref{fig:dq}, the black solid line clears that for $\lambda=0$ which corresponds to the standard canonical inflation, the prediction of the model lies outside the allowed regions of the Planck 2015 data. But in the non-canonical setting $(\lambda\neq0)$, figure shows that for some ranges of the parameter $\lambda$, the predictions of the model for $N_{*}=$ 50 and 60 can lie inside the 95\% CL region of the Planck 2015 data and for $N_{*}=40$ can enter the 68\% CL region. Using the Planck observational constraints on $r-n_{s}$ plane, we determine the allowed ranges of the parameter $\lambda$ for which the model (\ref{v:exp}) with the coupling term (\ref{w:exp}) is in consistency with the Planck 2015 data. The results are summarized in Table \ref{tab:tab1}. Also, using the $\lambda$ values in Table \ref{tab:tab1} we evaluate the running of the scalar spectral index $\frac{d n_{s}}{d\ln k}$ and the equilateral non-Gaussianity parameter $f_{\rm NL}^{\rm equil}$ with the help of Eqs. (\ref{dns:f:exp}) and (\ref{fNL:f:exp}), respectively. Table \ref{tab:tab1} shows that the values of $\frac{d n_{s}}{d\ln k}$ and $f_{\rm NL}^{\rm equil}$ predicted by our model are compatible with the Planck 2015 results.
\\
\\
\begin{table*}
        \caption{The $r-n_s$ consistency for the allowed ranges of the parameter $\lambda \equiv \mu/\alpha$ as well as the predicted values for the running of the scalar spectral index $\frac{d n_{s}}{d\ln k}$ and the equilateral non-Gaussianity parameter $f_{\rm NL}^{\rm equil}$ in the  model described by the exponential potential $V(\phi)=V_{0}\,e^{-\alpha \,(\phi/M_{\rm pl})}$ with the dilatonic coupling $\omega(\phi)= e^{-\mu (\phi/M_{\rm pl})}$.}
        \label{tab:tab1}
	    \begin{center}
            \scalebox{0.72}{
  \begin{tabular}{|>{\centering\arraybackslash}p{3.2cm}|>{\centering\arraybackslash}p{4.7cm}|
    >{\centering\arraybackslash}p{4.7cm} | >{\centering\arraybackslash}p{4.7cm}|
    >{\centering\arraybackslash}p{4.7cm}|}
    \hline

     \multirow{2}{*}{\diagbox[dir=SW,width=8.4em,height=5.4em,trim=r]{$V ={V_0}\,{e^ {-\alpha x}}$}{$\omega =e^ {-\mu x}$}}  &\multicolumn{2}{c|}{$N_* = 40$} &\multicolumn{1}{c|}{$N_* = 50$}&\multicolumn{1}{c|}{$N_* = 60$}
   \\

    \cline{2-5}
     & 68\% CL & 95\% CL    & 95\% CL &  95\% CL \\
      \hline
   $\lambda \equiv {\mu / \alpha}$ &  $\lambda \gtrsim {2.7} $ & $1.7 \lesssim \lambda \lesssim {2.7}$ & $\lambda \gtrsim {1.2}$  & $0.9 \lesssim \lambda \lesssim {10.9}$\\
      \hline
  $ \frac{ d{n_s}} {d\ln k}$ & $ {-0.0008} \lesssim \frac{d{n_s}}{d\ln k} \lesssim {-0.0006} $ & $ {-0.0010} \lesssim \frac{d{n_s}}{d\ln k} \lesssim {-0.0008} $ & $ {-0.0007} \lesssim \frac{d{n_s}}{d\ln k} \lesssim {-0.0004}$ &   $ {-0.0006} \lesssim \frac{d{n_s}}{d\ln k} \lesssim {-0.0003}$ \\
            \hline
           $f_{{\rm{NL}}}^{{\rm{equil}}}$  & $0.0104 \lesssim f_{{\rm{NL}}}^{{\rm{equil}}} \lesssim {0.0174}$ & $0.0174 \lesssim f_{{\rm{NL}}}^{{\rm{equil}}} \lesssim {0.0215}$  &  $0.0083\lesssim f_{{\rm{NL}}}^{{\rm{equil}}} \lesssim {0.0209}$   &   $0.0081 \lesssim f_{{\rm{NL}}}^{{\rm{equil}}} \lesssim {0.0202}$ \\
\hline
\end{tabular}
}
	\end{center}
\end{table*}

\section{inverse power-law potential in non-canonical inflation}\label{sec:inverse}

In this section, we focus on the inverse power-law potential which arises in supersymmetric theories and has the following form
\begin{equation}\label{invers:v}
V(\phi)=V_{0}\,{\left(\frac{\phi}{M_{\rm pl}}\right)^{-p}},
\end{equation}
where $V_{0}$ and $p>0$ are two constants of the model. This potential in the standard canonical setting leads to the intermediate inflation with the scale factor $a(t)\propto \exp [A({M_{\rm pl}}t)^f]$ where $A>0$, $0<f<1$ and $p=4(1-f)/f$ \cite{Bar90,Bar93, Bar06, Bar07, Mar14} and its prediction lies completely outside the region allowed by the Planck 2015 data \cite{Rezazade15,Rezazade2015}. It is also important to note that in the canonical scenario, the intermediate inflation like the power-law one has a graceful exit problem \cite{Bar93}. Therefore, it is interesting to see whether in the non-canonical framework, the potential (\ref{invers:v}) can be compatible with the Planck 2015 observations and the problem of ending inflation can be avoided. To do so, we consider the inverse power-law coupling as \cite{Farajollahi,Barrow}
\begin{equation}\label{invers:w}
\omega(\phi)=\left(\frac{\phi}{M_{\rm pl}}\right)^{-m},
\end{equation}
where $m$ is a constant. This kind of non-canonical coupling term is motivated by string theory. By applying the same approach that was followed for the exponential potential in the previous section, we can obtain the relations of $n_{s}$ and $r$ for the model considered here. To this end, using Eqs. (\ref{invers:v}) and (\ref{invers:w}) in (\ref{ns:V}) and (\ref{r:V}), we obtain $n_{s}$ and $r$ in terms of the scalar field $x$ as
\begin{align}
\label{ns:inv}
  n_{s}-1 &\simeq p\,\left(2-m-p\right)x^{-2+m} ,\\
  \label{r:inv}
  r & \simeq 8\,p^2 x^{-2+m}.
\end{align}
From Eq. (\ref{N:v}), the $e$-fold number reads
\begin{equation}\label{N:inv}
N\simeq\frac{1}{p\,(-2+m)}\left(x^{2-m}-x_{f}^{2-m}\right).
\end{equation}
Setting the condition of ending inflation $\varepsilon=1$ in Eq. (\ref{epsilon:v}), we can find
\begin{equation}\label{xf:inv}
x_{f}=\left(\frac{2}{p^2}\right)^{\frac{1}{-2+m}}.
\end{equation}
This clears that in contrary to the standard canonical scenario, in the non-canonical framework with the field-dependent coupling (\ref{invers:w}), a graceful exit can be provided for the inflationary potential (\ref{invers:v}).

Inserting Eq. (\ref{xf:inv}) into (\ref{N:inv}), we get
\begin{equation}\label{N:finv}
N\simeq\frac{1}{p\,(-2+m)}\left(x^{2-m}-\frac{p^2}{2}\right).
\end{equation}
Solving Eq. (\ref{N:finv}) for $x$, we obtain
\begin{equation}\label{x:inv}
x\simeq\left(N\,(-2+m)\,p+\frac{p^2}{2}\right)^{\frac{1}{2-m}}.
\end{equation}
Finally, replacing Eq. (\ref{x:inv}) into (\ref{ns:inv}) and (\ref{r:inv}) we can rewrite $n_{s}$ and $r$ in the following forms
\begin{align}
\label{ns:finv}
  n_{s}-1&\simeq \frac{2-m-p}{(\frac{p}{2})+ (-2+m)\,N}\,, \\
  \label{r:finv}
  r &\simeq\frac{8\,p}{(\frac{p}{2})+ (-2+m)\,N}\,.
\end{align}
Furthermore, using Eqs. (\ref{invers:v}) and (\ref{invers:w}) in (\ref{dns:V}) and (\ref{fNL:V}), one can rewrite the running of the scalar spectral index $\frac{d n_{s}}{d\ln k}$ and the equilateral non-Gaussianity $f_{\rm NL}^{\rm equil}$ as
   \begin{align}
  \label{dns:x:inv}
 \frac{d n_{s}}{d\ln k} &\simeq - (-2 + m)(-2 + m + p)~p^2~x^{-4 + 2 m},  \\
  \label{fNL:x:inv}
 f_{\rm NL}^{\rm equil} & \simeq \frac{5}{72}(-12 + 6 m + 11 p)~p~x^{-2 + m}.
\end{align}
With the help of Eq. (\ref{x:inv}), the above equations read
       \begin{align}
  \label{dns:f:inv}
 \frac{d n_{s}}{d\ln k} &\simeq -\frac{(-2 + m)(-2 + m + p)}{\Big((-2 + m)\,N + \frac{p}{2}\Big)^2}\,,  \\
  \label{fNL:f:inv}
 f_{\rm NL}^{\rm equil} & \simeq \frac{5}{72}\, \left(\frac{-12 + 6 m + 11 p}{ (-2 + m) N  + \frac{p}{2}}\right).
\end{align}
Now we turn to study the consistency of our model in the $r-n_s$ plane. Combining Eqs. (\ref{ns:finv}) and (\ref{r:finv}) leads to the linear relation
\begin{equation}\label{rns:inv}
r\simeq\left(\frac{8p}{2-m-p}\right)(n_{s}-1).
\end{equation}
When $m=0$, Eq. (\ref{rns:inv}) leads to the relation obtained for the inverse power-law potential in the standard canonical inflation as \cite{Bar06}
\begin{equation}\label{standard:inv}
n_{s}\simeq 1-\left(\frac{p-2}{8\,p}\right)r.
\end{equation}
\begin{figure}[t]
\begin{center}
\scalebox{0.95}[0.95]{\includegraphics{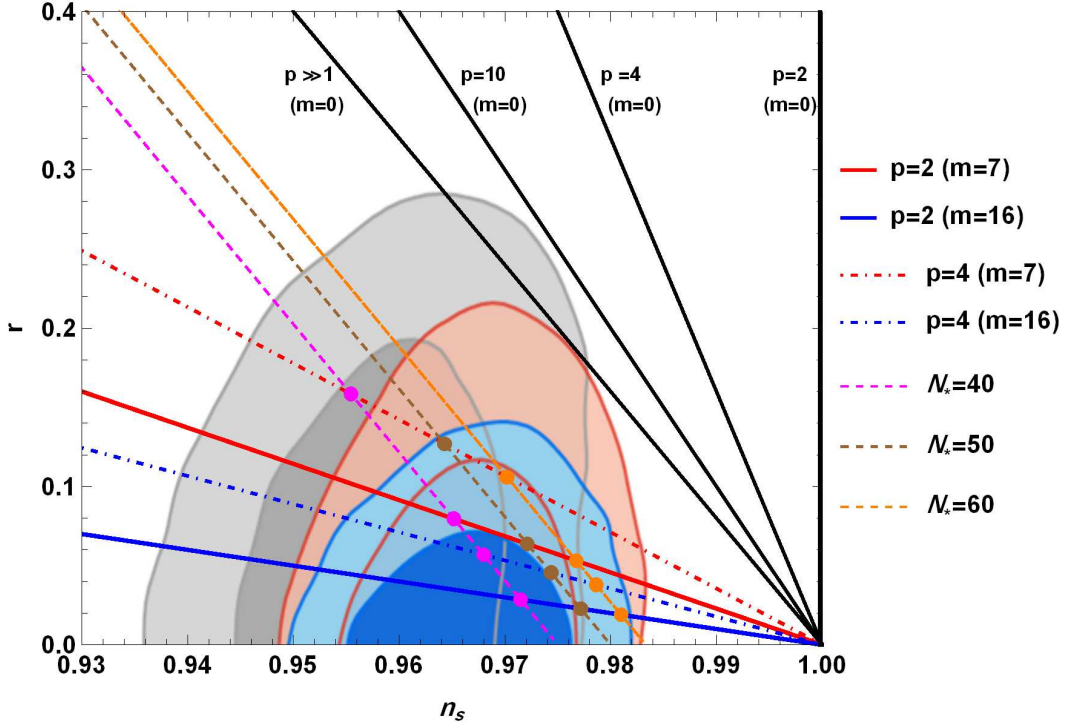}}
\caption{The $r-n_{s}$ diagram for the inverse power-law potential $V(\phi)=V_{0}\,{ (\phi/M_{\rm pl})^{-p}}$ with the field-dependent coupling $\omega(\phi)= (\phi/M_{\rm pl})^{-m}$. The results for $N_{*}=$ 40, 50 and 60 are shown by the pink, brown and orange dashed lines, respectively. Also, the predictions of the model for the two cases $p=2$ and 4 with $m=(7,16)$ are illustrated by the solid and dash-dotted color lines. Furthermore, the results for the standard canonical inflation ($m=0$) with different values of $p\geq 2$ are presented by the black solid lines.  The marginalized joint 68\% and 95\% CL regions of Planck 2013, Planck 2015 TT+lowP and Planck 2015
TT,TE,EE+lowP data \cite{Planck2015} are specified by gray, red and blue, respectively.}
\label{fig:inve1}
\end{center}
\end{figure}
\begin{figure}[t]
\begin{center}
\scalebox{0.95}[0.95]{\includegraphics{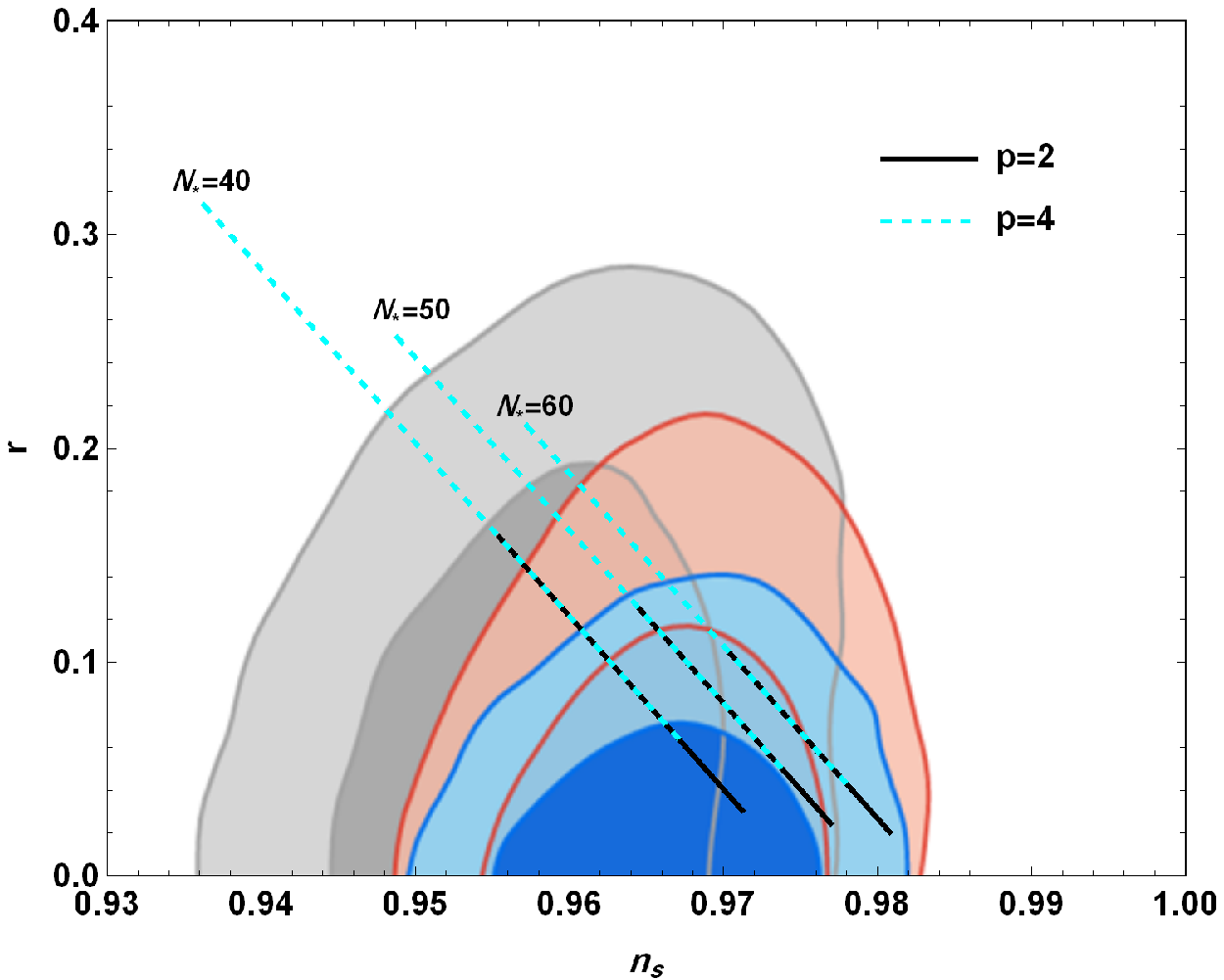}}
\caption{The $r-n_{s}$ diagram for the inverse power-law potential $V(\phi)=V_{0}\,{ (\phi/M_{\rm pl})^{-p}}$ with the field-dependent coupling $\omega(\phi)= (\phi/M_{\rm pl})^{-m}$. The results for $p=2$ and $p=4$ with $N_{*}=(40,50,60)$ are shown by the black solid and color dashed lines, respectively. Here $4.5\leq m\leq 15$. The marginalized joint 68\% and 95\% CL regions of Planck 2013, Planck 2015 TT+lowP and Planck 2015
TT,TE,EE+lowP data \cite{Planck2015} are specified by gray, red and blue, respectively.}
\label{fig:N}
\end{center}
\end{figure}
The above equation clearly indicates that for $p<2$ since $r>0$ then we have $n_{s}>1$ which is outside the 95\% CL region of the Planck 2015 data. For $m=0$, using Eq. (\ref{standard:inv}) the $r-n_{s}$ diagram for different values of $p\geq 2$ are shown in Fig. \ref{fig:inve1}. The figure illustrates that the predictions of the inverse power-law potential (\ref{invers:v}) in the standard canonical framework ($m=0$) for $p\geq 2$, like the case $p<2$ as discussed above, are not consistent with the Planck 2015 data. Also for $m=0$ in the limit of $p\gg1$, Eq. (\ref{standard:inv}) reduces to the relation $r=8\,(1-n_{s})$. This shows that the result of intermediate inflation (i.e. $a(t)\propto \exp [A({M_{\rm pl}}t)^f]$ where $A>0$, $0<f<1$ and $p=4(1-f)/f$) in canonical setting for $p\gg1$, as shown in Fig. \ref{fig:inve1}, is same as that obtained for the power-law inflation (i.e. $a(t)\propto t^{q}$ where $q>1$) in the standard canonical scenario \cite{Mar14}. See also the black solid line in Fig. \ref{fig:dq}.

In Fig. \ref{fig:inve1}, using Eqs. (\ref{ns:finv}) and (\ref{r:finv}) we further plot the $r-n_s$ diagram for the non-canonical scenario ($m\neq0$). The pink, brown and orange dashed lines are corresponding to $N_{*}=$ 40, 50 and 60, respectively, with varying $p$ and $m$. Our investigation shows that for a given $N_*$, the results for different $p$ values with varying $m$ overlap with each other. Figure \ref{fig:N} shows this in more details. For instance, in Fig. \ref{fig:N}, for $N_*=40$ when $4.5\leq m \leq 15$ the result for $p=2$ overlaps with that one obtained for $p=4$. This can also be understood in another way. With the help of Eqs. (\ref{ns:finv}) and (\ref{r:finv}) one can find
\begin{equation}
\frac{dr}{dn_s}\Big|_{N=N_*={\rm cte.}}=-\left(\frac{16N_*}{2N_*-1}\right),
\end{equation}
which shows that the slope of $r-n_s$ diagram when $N=N_*={\rm cte.}$, is independent of $p$ and $m$ values.

In Fig. \ref{fig:inve1}, using Eq. (\ref{rns:inv}) we also plot the predictions of the model for the two cases $p=2$ and 4 with different values of $m=(7,16)$. From the figure, we see that in the non-canonical framework ($m\neq 0$) for $N_{*}=50$ and $N_{*}=60$, the results of the model (for any value of $p>0$) can lie within the 95\%CL region allowed by Planck 2015 TT, TE, EE+lowP data and if we take $N_{*}=40$, the results can enter the 68\% CL region.
In Table \ref{tab:tab2}, using the $r-n_s$ constraint on the model parameters we present the allowed ranges of the parameter $m$ for different values of $p=1,2,3,4,5,6$. Besides, Table \ref{tab:tab2} summarizes the predictions of the model for $d{n_s}/d\ln k$ and $f_{\rm NL}^{\rm equil}$ obtained using (\ref{dns:f:inv}) and (\ref{fNL:f:inv}), respectively. The results for $d{n_s}/d\ln k$ and $f_{\rm NL}^{\rm equil}$ are consistent with the Planck 2015 observations. Note that the values of $d{n_s}/d\ln k$ and $f_{\rm NL}^{\rm equil}$ listed in Table \ref{tab:tab2} are same as those obtained in Table \ref{tab:tab1}. This is because of Eqs. (\ref{dns:f:inv}) and (\ref{fNL:f:inv}) by changing variable $\frac{m-2}{p}\rightarrow \lambda$ reduce to Eqs. (\ref{dns:f:exp}) and (\ref{fNL:f:exp}), respectively.
\begin{table*}
        \caption{The $r-n_s$ consistency for the allowed ranges of the parameter $m$ for different values of $p$ as well as the predicted values for the running of the scalar spectral index $\frac{d n_{s}}{d\ln k}$ and the equilateral non-Gaussianity parameter $f_{\rm NL}^{\rm equil}$ in the  model described by the inverse power-law potential $V(\phi)=V_{0}\,{ (\phi/M_{\rm pl})^{-p}}$ with the filed-dependent coupling $\omega(\phi)=(\phi/M_{\rm pl})^{-m}$.}
        \label{tab:tab2}
	\begin{center}
\scalebox{0.72}{
  \begin{tabular}{|>{\centering\arraybackslash}p{3.2cm}|>{\centering\arraybackslash}p{4.7cm}|
    >{\centering\arraybackslash}p{4.7cm} | >{\centering\arraybackslash}p{4.7cm}|
    >{\centering\arraybackslash}p{4.7cm}|}
    \hline

     \multirow{2}{*}{\diagbox[dir=SW,width=8.4em,height=5.4em,trim=r]{$V ={V_0}\,{x^{-p}}$}{$\omega =x^{-m}$}}  &\multicolumn{2}{c|}{$N_* = 40$} &\multicolumn{1}{c|}{$N_* = 50$}&\multicolumn{1}{c|}{$N_* = 60$}
   \\

    \cline{2-5}
     & 68\% CL & 95\% CL    & 95\% CL &  95\% CL \\
     \hline
            $p=1$ &  $ m\gtrsim {4.7} $ & $ 3.7 \lesssim m \lesssim {4.7}$ &$ m \gtrsim {3.2}$ & $ 2.9 \lesssim m \lesssim {12.9}$\\
      \hline

   $p=2$ &  $ m \gtrsim {7.4} $ & $ 5.4\lesssim m \lesssim {7.4}$ & $ m \gtrsim {4.4}$ & $ 3.9 \lesssim m \lesssim {23.8}$\\
      \hline

  $p=3$ &  $ m \gtrsim {10.1} $ & $ 7.1 \lesssim m \lesssim {10.1}$ & $ m \gtrsim {5.6}$ & $ 4.8 \lesssim m \lesssim {34.7}$\\
      \hline

  $p=4$ &  $ m \gtrsim {12.8} $ & $ 8.8 \lesssim m \lesssim {12.8}$ & $ m \gtrsim {6.8}$  & $ 5.8 \lesssim m \lesssim {45.6}$\\
       \hline

   $p=5$ &  $ m \gtrsim {15.5} $ & $ 10.5 \lesssim m \lesssim {15.5}$  & $ m \gtrsim {8.0}$ & $ 6.7 \lesssim m \lesssim {56.5}$\\
      \hline

    $p=6$ &  $ m \gtrsim {18.2} $ & $ 12.2 \lesssim m \lesssim {18.2}$ & $ m \gtrsim {9.2}$ & $ 7.7 \lesssim m \lesssim {67.4}$\\
      \hline

   $ \frac{d n_s}{d\ln k}$ & $ {-0.0008} \lesssim \frac{d{n_s}}{d\ln k} \lesssim {-0.0006} $ & $ {-0.0010} \lesssim \frac{d{n_s}}{d\ln k} \lesssim {-0.0008} $ & $ {-0.0007} \lesssim \frac{d{n_s}}{d\ln k} \lesssim {-0.0004}$ & $ {-0.0006} \lesssim \frac{d{n_s}}{d\ln k} \lesssim {-0.0003}$ \\
      \hline

   $f_{{\rm{NL}}}^{{\rm{equil}}}$  & $0.0104 \lesssim f_{{\rm{NL}}}^{{\rm{equil}}} \lesssim {0.0174}$ & $0.0174 \lesssim f_{{\rm{NL}}}^{{\rm{equil}}} \lesssim {0.0215}$   & $0.0083\lesssim f_{{\rm{NL}}}^{{\rm{equil}}} \lesssim {0.0209}$   & $0.0081 \lesssim f_{{\rm{NL}}}^{{\rm{equil}}} \lesssim {0.0202}$ \\
\hline
\end{tabular}
}
	\end{center}
\end{table*}

\section{Conclusions}\label{sec:con}

Here, we investigated inflation within the framework of non-canonical scalar field. With the help of scalar and tensor spectrum, we first obtained the basic relations governing the inflationary observables including the scalar spectral index $n_{s}$, the tensor-to-scalar ratio $r$, the running of the scalar spectral index $d{n_s}/d\ln k$ and the equilateral non-Gaussianity parameter $f_{\rm NL}^{\rm equil}$ in terms of the potential $V(\phi)$ and the non-canonical coupling $\omega(\phi)$. Then, we applied these results for the two models containing the exponential potential $V(\phi)=V_{0}\,{e^{-\alpha (\phi/M_{\rm pl})}}$ with the dilatonic coupling $\omega(\phi)=e^{-\mu (\phi/M_{\rm pl})}$ and the inverse power-law potential $V(\phi)=V_{0}(\phi/M_{\rm pl})^{-p}$ with the field coupling $\omega(\phi)=(\phi/M_{\rm pl})^{-m}$. We found that in contrary to the standard canonical inflation (i.e. $\omega(\phi)=1$), the predictions of both models in the $r-n_{s}$ plane for $N_{*} = 50$ and $N_{*} = 60$ can lie inside the 95\% CL region of the Planck 2015 TT,TE,EE+lowP data and for $N_{*} = 40$ can enter the 68\% CL region. Using the $r-n_s$ constraint, we also determined the allowed ranges of the constant parameters appeared in the both models. Finally, we estimated the running of the scalar spectral index $d{n_s}/d\ln k$ and the equilateral non-Gaussianity parameter $f_{\rm NL}^{\rm equil}$ for the both models and conclude that the results of these two observables are also consistent with the Planck 2015 observations. It is also worth to mentioning that although the exponential and inverse power-law potentials in the standard canonical setting suffer from the graceful exit problem in which inflation never ends, this central drawback is resolved in non-canonical inflation.



\end{document}